\documentclass[sigconf]{acmart}
\acmConference[ASE 2022]{37th IEEE/ACM International Conference on Automated Software Engineering}{Mon 10 - Fri 14 October 2022}{Oakland Center}

\usepackage{listings}
\usepackage{balance}
\usepackage{caption}
\usepackage{subcaption}
\usepackage{tikz}
\usetikzlibrary{tikzmark}
\usepackage{color}
\usepackage{enumitem}

\definecolor{mygreen}{rgb}{0,0.6,0}
\definecolor{mygray}{rgb}{0.5,0.5,0.5}
\definecolor{mymauve}{rgb}{0.58,0,0.82}

\newcommand{\mytikzmark}[1]{%
  \tikz[overlay,remember picture,baseline] \coordinate (#1) at (0,0) {};}

\usepackage{diagbox}
\usepackage{tabularx}
\usepackage{booktabs}
\usepackage{semantic}
\usepackage{proof}
\usepackage{numprint}
\npdecimalsign{\ensuremath{\cdot}}
\npthousandsep{,}

\usepackage{soul}

\def\*#1{\mathbf{#1}}

\AtBeginDocument{%
  \providecommand\BibTeX{{%
    \normalfont B\kern-0.5em{\scshape i\kern-0.25em b}\kern-0.8em\TeX}}}

\copyrightyear{2022}
\acmYear{2022}
\setcopyright{rightsretained}
\acmConference[ASE '22]{37th IEEE/ACM International Conference on Automated Software Engineering}{October 10--14, 2022}{Rochester, MI, USA}
\acmBooktitle{37th IEEE/ACM International Conference on Automated Software Engineering (ASE '22), October 10--14, 2022, Rochester, MI, USA}
\acmDOI{10.1145/3551349.3556918}
\acmISBN{978-1-4503-9475-8/22/10}

\begin{document}

\hyphenation{in-for-ma-tion}

\title{Data Leakage in Notebooks: Static Detection and Better Processes}


\author{Chenyang Yang}
\affiliation{%
  \institution{Carnegie Mellon University}
  \country{}}
  
\author{Rachel A Brower-Sinning }
\affiliation{%
  \institution{Carnegie Mellon Software Engineering Institute}
  \country{}}
  
 \author{Grace A. Lewis}
\affiliation{%
  \institution{Carnegie Mellon Software Engineering Institute}
  \country{}}
  
\author{Christian K\"astner}
\affiliation{%
  \institution{Carnegie Mellon University}
  \country{}}

\renewcommand{\shortauthors}{Yang, et al.}

\begin{abstract}
Data science pipelines to train and evaluate models with machine learning may contain bugs just like any other code.
Leakage between training and test data can lead to overestimating the model's accuracy during offline evaluations, possibly leading to deployment of low-quality models in production.
Such leakage can happen easily by mistake or by following poor practices, but may be tedious and challenging to detect manually.
We develop a static analysis approach to detect common forms of data leakage in data science code.
Our evaluation shows that our analysis accurately detects data leakage and that such leakage is pervasive among over \numprint{100000} analyzed public notebooks.
We discuss how our static analysis approach can help both practitioners and educators, and how leakage prevention can be designed into the development process.
\end{abstract}





\maketitle


\section{Introduction}
\label{sec:intro}

\begin{figure}
\begin{lstlisting}[language=Python] 
import numpy as np
# generate random data
n_samples, n_features, n_classes = 200, 10000, 2
rng = np.random.RandomState(42)
X = rng.standard_normal((n_samples, n_features))
y = rng.choice(n_classes, n_samples)

# leak test data through feature selection
X_selected = SelectKBest(k=25).fit_transform(X, y)

X_train, X_test, y_train, y_test = train_test_split(
    X_selected, y, random_state=42)
gbc = GradientBoostingClassifier(random_state=1)
gbc.fit(X_train, y_train)

y_pred = gbc.predict(X_test)
accuracy_score(y_test, y_pred)  
# expected accuracy ~0.5; reported accuracy 0.76
\end{lstlisting}
\vspace{-1em}
\caption{Data leakage may cause a highly-biased test result.
The model learns test data distribution through feature selection,
resulting in an over-optimistic test score.
}
\label{fig:leak_acc}
\vspace{-1.5em}
\end{figure}

Will a promising machine-learned model work when deployed in production? 
Typically this question is answered by comparing model predictions to expected outcomes on test data.
However, the resulting accuracy estimates can be misleading, where the model performs well on test data, but much worse in production. 
A common cause is that the data used for testing is not representative enough of the production data, thus providing misleading estimates on the wrong data distribution. 
A different cause, and the focus of this paper, is that the test data was used in some form during model training (directly or indirectly, intentionally or accidentally) allowing the model to overfit on the test data, thus producing unrealistically optimistic accuracy estimates. Because data science pipelines are code, we can use software engineering techniques to analyze them---which we do in this paper.

In this paper, we design a static analysis approach to detect cases where model training makes use of test data in data science code, commonly called \emph{data leakage}~\cite{kaufman2012leakage, burkov2020machine}. Data leakage is often the result of using bad practices when writing machine learning code, ranging from obvious mistakes, such as including test data in the training data, to more subtle ones that leak test data distribution information through preprocessing prior to training. For example, in Fig.~\ref{fig:leak_acc}, we show data science code reporting confidently to find patterns in random data where the model should not do better than a random guess: Because decisions during training depend on both training and test data (feature selection, Line~9) the model overfits on test data and the evaluation reports significantly inflated accuracy scores. Our analysis points out common pitfalls in model accuracy evaluations like the one in our example, which, as we will show, are pervasive in data science code in public notebooks.

Our analysis has both practical and educational value. On the practical side, our work contributes to more reliable offline evaluations of machine-learned models, which are an important quality assurance step when integrating models into software products. Use of machine learning in software products is increasingly common, but also very challenging~\cite{nadiaICSE22,microsoftSE4MLpaper,hulten2018building}. Reliable offline accuracy evaluations are important for preventing harm from deploying low-quality models in production systems, where harm can range from stress, to discrimination, to fatal accidents~\cite{amazon-bias, tesla-crash}.
Accuracy results are also a common quality metric between teams~\cite{nadiaICSE22}, especially when delegating or entirely outsourcing model development. When accuracy goals are parts of contracts (or competitions) there may be an incentive to report inflated accuracy results.

\looseness=-1
On the educational side, the danger of overfitting and data leakage is well known and commonly discussed in textbooks~\cite{MLtextbook, burkov2020machine}, ML library documentation~\cite{scikit, imblearn}, and tutorials~\cite{alexisbcook_2021}. Yet, as we will show, leakage also occurs in tutorial notebooks, popular notebooks, and entries in data science competitions, which others may use as educational resources or templates. Our analysis, just like other static analyses, can help raise awareness of coding problems and nudge students and model developers toward better practices.

Technically, we develop a static data-flow analysis that tracks how datasets flow through data science code and are used in training and evaluation functions of machine-learning libraries. To allow accurate detection, we track specific kinds of transformations and detect common patterns that lead to leakage. In an evaluation with data science code from public notebooks, we show that our analysis is accurate (92.9\%) with very few false positives and can analyze most notebooks within a few seconds. Applying our analysis to over 100,000 public notebooks, we detect data leakage issues in nearly 30~percent of them.

In summary, we make the following contributions:
\begin{itemize}
    \item  A summary and formulation of common data leakage problems.
\item A static analysis that can automatically detect data leakage.
\item Results from a large-scale study on data leakage in public notebooks.
\item Recommendations on process designs that prevent data leakage.
\end{itemize}
    
We share our tool and supplementary materials on GitHub.\footnote{https://github.com/malusamayo/leakage-analysis}

\section{Overfitting and Data Leakage in Machine Learning}\label{sec:overfitting}

Machine learning is the discipline of learning generalizable insights
from data, typically in the form of a learned function, called \textit{model}, that can make predictions
for unseen data (e.g., production data). Developers building models with machine learning techniques
usually follow an iterative and exploratory process~\cite{kery2018story} that is
commonly depicted as a pipeline of multiple steps with feedback loops, including activities such as
data collection, data cleaning, feature engineering, model training,
 model evaluation, and model deployment~\cite{microsoftSE4MLpaper}.
 
 
In model development, there is always the risk that the trained model \textit{overfits} on
the data used for training~\cite{Russell1995ArtificialIA}---that is, it learns the patterns in the specific training
data but generalizes poorly to unseen data.
Therefore, it is customary to evaluate the accuracy of a model on 
data that was not previously used for training~\cite{Russell1995ArtificialIA}---the evaluation measures
to what degree the model predicts expected results for unseen data.
For the evaluation to provide a meaningful approximation of the model's accuracy in production settings, the unseen data needs to be representative of the distribution of real data encountered in production.

Overfitting can happen whenever insight is gained from data,
whether it is (a) a machine learning algorithm that is learning model parameters from
data or (b) a human looking at data to make decisions about how to process the data
or about what machine learning algorithm to use.
Most importantly, due to the iterative nature of model development, it is common to evaluate
 different variants of a model to see whether accuracy improves
with different decisions (e.g., different feature engineering, different
machine-learning algorithm, different hyperparameters; some of this
exploration may also be automated using AutoML approaches~\cite{He2021AutoMLAS}). If decisions are based on prior evaluation,
the data used in that evaluation influenced the training process and
the model may overfit on it.

In summary, if we evaluate the model on data that was used in any form 
(automated or manually, directly or indirectly) in the socio-technical process used for training the model, the 
evaluation result may be overly optimistic because the model may have overfit
on that data.
In a technical sense, we want a \textit{non-interference guarantee} in
which the process of training the model is entirely
independent of the data on which the model is evaluated.

\paragraph{Offline/Online Evaluation}
The model evaluation we discuss \linebreak above is usually executed offline \textit{before} model deployment.
Model developers could also conduct an online evaluation with production data \textit{after} their model is deployed. 
Typically offline evaluations are conducted to gain confidence in the model before deployment and to avoid exposing users to low-quality models in production, 
just like software developers rely on unit testing to identify software bugs rather than only relying on crash reports and bug reports from users in production.


\paragraph{Training-Validation-Test Splits}

In many settings, labeled data that can be used for training or evaluation is limited and expensive to gather. Many data science projects start with a single dataset, from which separate subsets are used for training and evaluation.
The most common approach is to split data three ways into 
\textit{training data}, \textit{validation data}, and \textit{test data}.
Training data is used to develop the model and validation data is used
for preliminary evaluation during model development (including hyperparameter tuning), whereas test data should just be used once as a final unbiased evaluation of the final model.
Validation and test data seem similar and they are often used in the same type of evaluation functions in  machine learning APIs, but they serve fundamentally different purposes---validation data is used for decision making during model development and hence not suitable for an independent evaluation. 

The concepts of overfitting and the need to properly split data into these three sets to achieve unbiased evaluation results are universally covered in machine learning education and explained extensively in textbooks and course materials~\cite[e.g.,][]{Russell1995ArtificialIA, MLtextbook}.





\begin{figure}


\begin{subfigure}[b]{\columnwidth}
\begin{lstlisting}[language=Python]
# oversampling datasets, new rows are synthesized based on existing rows
X_new,y_new = SMOTE().fit_resample(X,y)
# splits after over-sampling no longer produce independent train/test data
X_train, X_test, y_train, y_test = train_test_split(X_new, y_new, test_size=0.2, random_state=42)

rf = RandomForestClassifier().fit(X_train,y_train)
rf.predict(X_test)
\end{lstlisting}
\vspace{-1em}
\caption{\small Test data used for training}
\vspace{0.5em}
\label{code:overlap_example}
\end{subfigure}


\begin{subfigure}[b]{\columnwidth}
\begin{lstlisting}[language=Python]
# select the best model with repeated evaluation
results = []
for clf, name in (
        (DecisionTreeClassifier(), "Decision Tree"),
        (Perceptron(), "Perceptron")):
    clf.fit(X_train, y_train) 
    pred = clf.predict(X_test) 
    score = metrics.accuracy_score(y_test, pred)
    results.append(score, name)
\end{lstlisting}
\vspace{-1em}
\caption{\small Test data used repeatedly for model selection}
\vspace{0.5em}
\label{code:multi_test_example}
\end{subfigure}


\begin{subfigure}[b]{\columnwidth}
\begin{lstlisting}[language=Python]
# unknown words in test data leak into training data
wordsVectorizer = CountVectorizer().fit(text)
wordsVector = wordsVectorizer.transform(text)
invTransformer = TfidfTransformer().fit(wordsVector)
invFreqOfWords = invTransformer.transform(wordsVector)
X = pd.DataFrame(invFreqOfWords.toarray())

train, test, spamLabelTrain, spamLabelTest = train_test_split(X, y, test_size = 0.5)    
predictAndReport(train = train, test = test)
\end{lstlisting}
\vspace{-1em}
\caption{Test data leaked in preprocessing}
\label{code:pre_processing_example}
\end{subfigure}

\caption{Shortened data leakage examples from public notebooks.}
\label{fig:example}
\vspace{-2em}
\end{figure}


\paragraph{Data Leakage}
Despite the conceptual requirement to never make any decisions 
that influence the model based on data that is used for evaluating
the final model (i.e., noninterference of test data on model training), 
in practice, violations of this requirement are common and known as
\textit{data leakage} (because test data ``leaks'' into the training process)~\cite{burkov2020machine, kaufman2012leakage, DBLP:journals/corr/abs-2110-08339, deepchecks}. We target three forms of data leakage:


\begin{itemize}
    \item 
\textbf{Overlap Leakage:}
An obvious form of leakage occurs when some or all test data is directly used as input for training or hyper-parameter tuning. 
More subtly, leakage can occur when creating training data based on test data in the form of data augmentation or oversampling, as in Fig.~\ref{code:overlap_example}.
We call this type of leakage \textit{overlap leakage}, as rows of test data overlap with rows of training data. 

\item 
\textbf{Multi-Test Leakage:} 
If data is used repeatedly for evaluation, it is highly likely that
decisions are made based on that data, including algorithm selection, 
model selection, and hyperparameter tuning. For example, data scientists may have selected the model that works best on the data. Data used repeatedly in evaluation, as in Figure~\ref{code:multi_test_example},
can no longer be considered as unseen \textit{test} data, but should be considered as \textit{validation} data.

\item 
\textbf{Preprocessing Leakage:} 
When training data and test data are preprocessed (transformed) together, 
test data sometimes influences the transformations of the training data.
For example, data could be normalized according to the largest and smallest values in both training and test data, rather than only based on values from training data.
Preprocessing leakage can occur in many transformations that consider multiple rows of the dataset,  
including feature selection (e.g., Fig.~\ref{fig:leak_acc}), normalizing data, projecting data with PCA, and vectorizing text data (e.g., Fig.~\ref{code:pre_processing_example}).
In many practical settings, training and test data have very similar distributions and preprocessing leakage has  only marginal influence on training data and hence the model; however, it is easy to construct examples where the mere knowledge about the distribution of test data can lead to substantial overfitting (see Fig.~\ref{fig:leak_acc}) and
 out-of-distribution predictions are particularly affected. 
\end{itemize}

We target these forms of leakage because they are common sources of overconfident evaluation results, are discussed frequently both by practitioners and the literature~\cite{burkov2020machine, DBLP:journals/corr/abs-2110-08339}, and could be detected with static inspection of source code without understanding of the semantics of the data.
Other forms of leakage are beyond the scope of this paper, including \textit{label leakage} where unintended features in the data correlate with labels, leading to shortcut learning \cite{geirhos2020shortcut, kaufman2012leakage,burkov2020machine}; leakage from \textit{incorrect splits} of data when dependencies between rows exist, such as in time-series data~\cite{splits-aiops,burkov2020machine}---both of these forms of leakage require a deep understanding of the semantics of the data and are orthogonal to the three forms of data leakage we address.

    




\section{Approach}
\begin{figure}[t]
   \centering
  \includegraphics[width=\columnwidth]{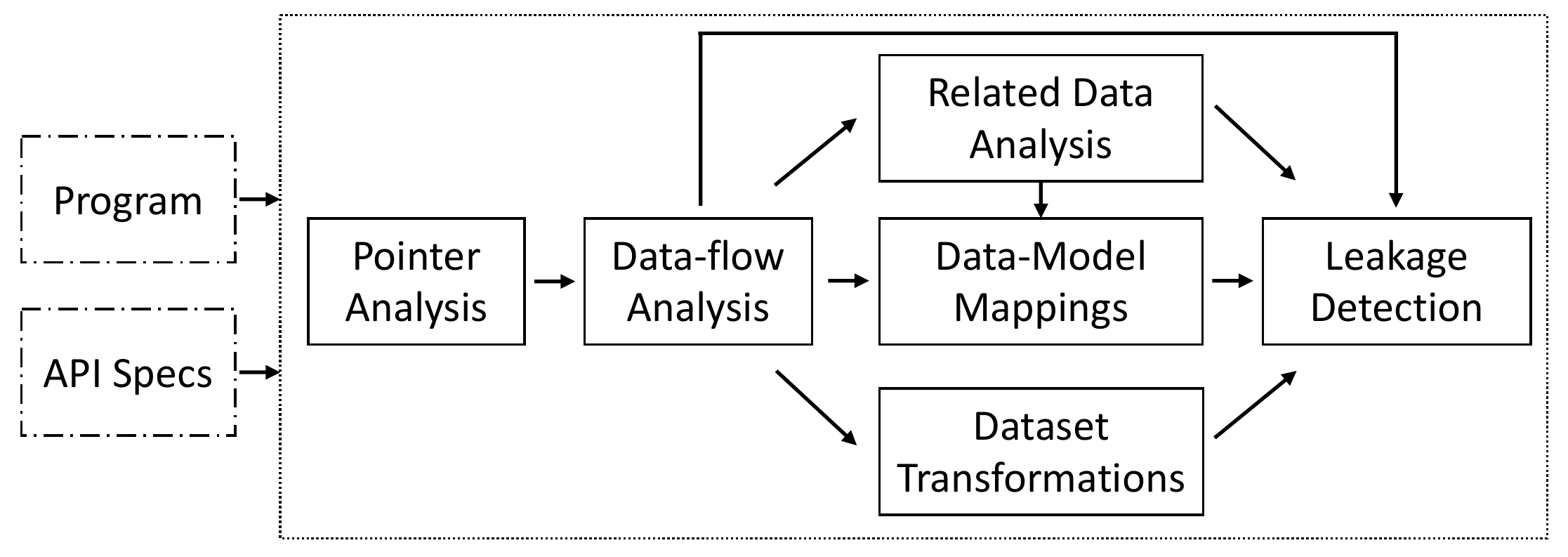}
  \caption{Approach Overview.
  Our analysis first performs standard pointer analysis and data-flow analysis,
  and collects domain-specific information (dataset transformations, related-data, data-model mappings) from the results.
  Finally, leakage is detected using all the information collected.
}
  \label{fig:approach}
  \vspace{-1em}
\end{figure}


\begin{figure*}[ht]
\includegraphics{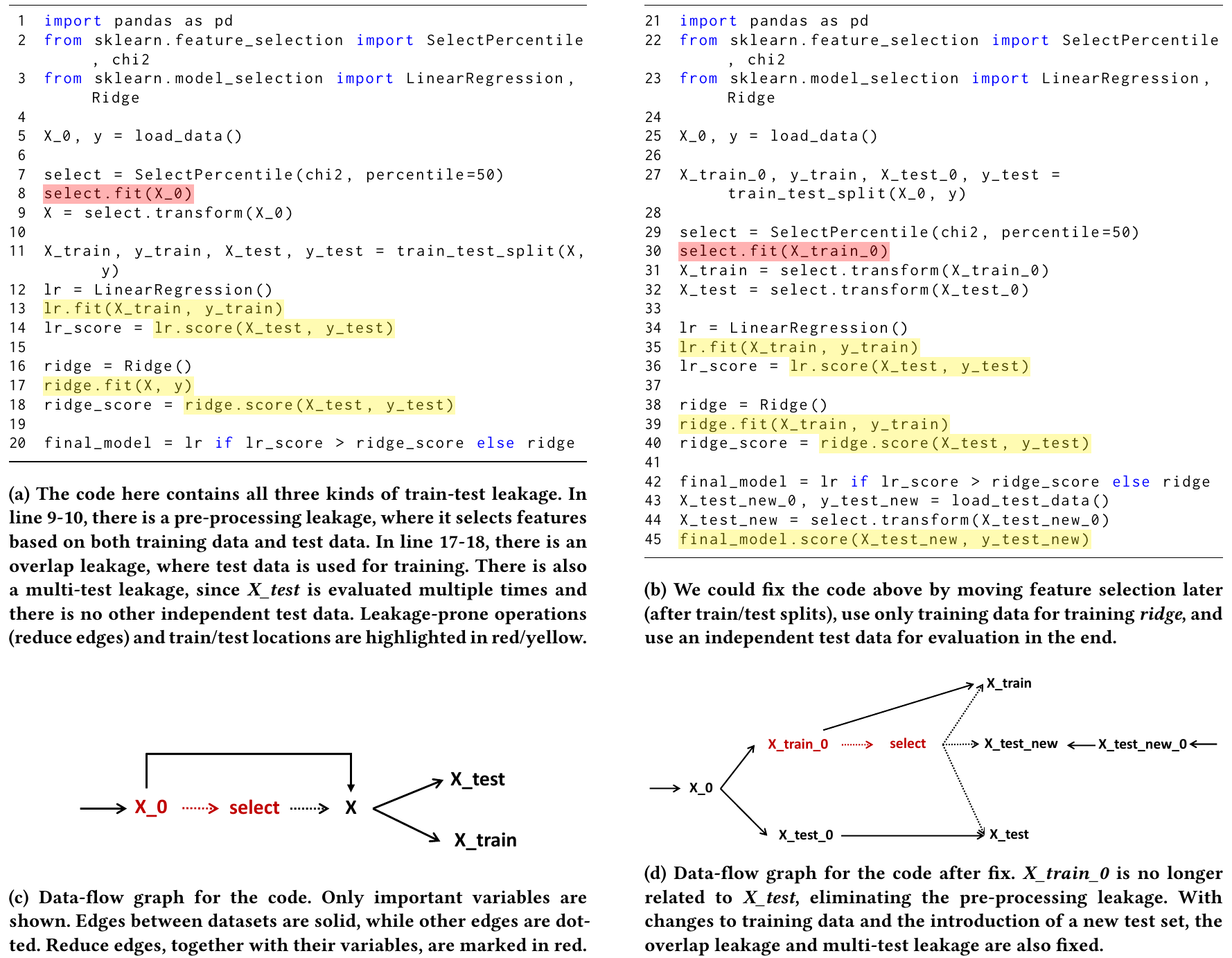}
\caption{Example code with data-flow graph (before and after fix) for leakage analysis.}
\label{fig:approach_example}
\end{figure*}


\noindent We developed a static code analysis approach to detect different forms of data leakage. 
We analyze how data flows through notebook code and how it is used for training, validation, and testing.
To this end, we statically collect specific information needed to detect leakage (see Fig.~\ref{fig:approach}):
\begin{itemize}\item
\textbf{Dataset transformations.} In the preprocessing steps, transformations may leak information across rows, duplicate rows, or transform rows independently. 
Some of these transformations could contribute to preprocessing leakage or overlap leakage. 
Therefore, we need to track how datasets are processed. To this end, we label data-flow edges to track different kinds of dataset transformations.

\end{itemize}
\begin{itemize}\item
\textbf{Related-data relations.} To map data to models and detect overlap leakage, we need to understand whether datasets may have originated from the same rows in an original dataset. 
We track this with a related-data\emph{ relation} on top of our standard data-flow relations that tracks which two variables are related.

\end{itemize}
\begin{itemize}\item
\textbf{Data-model mappings}. To detect leakage, we need to identify the training/validation/test data for a given model. 
Here the key challenge is to differentiate validation and test data. 
We collect this information in data-model mappings, built on top of the related-data relations.

\end{itemize}
Once we have collected data flows (including data transformations and related-data relations) and data-model mappings, we can detect leakage by matching patterns over this information. In the remainder of this section, we explain each of these steps using a running example (Fig.~\ref{fig:approach_example}).

\subsection{Tracking Data Flows}

For all leakage detection, we need to identify how datasets and other computations relate to each other in the notebook. 
That is, we need to track how data may be repeatedly transformed and split, possibly by using information originally derived from other parts of the data, until it flows into training or evaluation functions of models. 
To this end, we perform standard data-flow analysis through assignments and method calls. In addition, we collect additional information about dataset transformations and related-data relations:


\begin{figure}[t]
\small

\begin{tabular}{l}

\textbf{Inputs} \hfill \\
 $\mathbf{V}$ \hfill program variables  \\
 $\mathbf{D} \subseteq \mathbf{V}$ \hfill datasets  \\
 $\mathbf{M} \subseteq \mathbf{V}$ \hfill models  \\
 $\textit{DataFlow} \subseteq \*V\times \*V$ \hfill data flow paths (transitive closure)  \\
 $\textit{DatasetFlow} \subseteq \*D\times \*D$ \hfill data flow paths between datasets \\
 $\textit{MapEdge} \subseteq \*D\times \*D$ \hfill map-like operations \\
 $\textit{ReduceEdge} \subseteq \*D\times \*V$ \hfill reduce-like operations \\
 $\textit{DupEdge} \subseteq \*D\times \*D$ \hfill duplication operations \\
 $\textit{ModelData} \subseteq \*M\times \*D\times \mathcal{P}(\*D)\times  \mathcal{P}(\*D)\quad$ \hfill models with corresponding \\
 \hfill training, validation, and test datasets \\
 \\
 
\textbf{Rules} \hfill \\

 
\end{tabular}

\footnotesize
\[
    \begin{array}{c}  \\
      \inference{a \in \*D \quad b\in \*D\quad (a, b) \in \textit{DatasetFlow}}
      {(a, b) \in \textit{RelData}}[reldata/flow]
      \\[1.5em]
      \inference{(b, a) \in \textit{RelData}}
      {(a, b) \in \textit{RelData}}[reldata/sym] 
      \hspace{2em}\inference{a \in \*D}
      {(a, a) \in \textit{RelData}}[reldata/ref] 
      \\[1.5em]
      \inference{(a, b) \in \textit{RelData}\quad (a, c) \in \textit{RelData}\quad (a, c) \in \textit{MapEdge}}
      {(b, c) \in \textit{RelData}}[reldata/map]
      \\[1.5em]
      \inference{(src, a) \in \textit{DupEdge}\quad (a, b) \in \textit{RelData}\quad (a, c) \in \textit{RelData}}
      {(b, c) \in \textit{RelData}}[reldata/dup]
      \\[1.5em]
      \inference{\exists(m, d_\textit{tr}, D_\textit{va}, D_\textit{te})\in \textit{ModelData},\; D_\textit{va} \neq \varnothing\\ \forall (m, d_\textit{tr}, D_\textit{va}, D_\textit{te})\in \textit{ModelData},\; D_\textit{te} = \varnothing}
      {\textit{multi-test leakage in notebook}}[leak/multi] 
      \\[1.5em]
      \inference{(m, d_\textit{tr}, D_\textit{va}, D_\textit{te})\in \textit{ModelData}\\ \forall d_e\in D_\textit{te}\cup D_\textit{va}, (d_e, d_\textit{tr})\in \textit{RelData}}
      {\textit{leakage due to overlap between $d_\textit{tr}$ and $D_\textit{te}\cup D_\textit{va}$ for model $m$}}[leak/overlap]
      \\[1.5em]
      \inference{(m, d_\textit{tr}, D_\textit{va}, D_\textit{te})\in \textit{ModelData}\quad
      (s, t) \in \textit{ReduceEdge} \\ d_\textit{e}\in  D_\textit{te}\cup D_\textit{va}\quad (d_\textit{e}, s) \in \textit{RelData}\quad (t, d_\textit{tr})\in \textit{DataFlow} }
      {\textit{leakage between $d_\textit{tr}$ and $d_\textit{e}$ through the transformation from $s$ to $t$}}[leak/pre] 
    \end{array}
  \]
  \vspace{-1em}
  \caption{Notations and rules for leakage detection from input relations.}
  \label{fig:rules}
\end{figure}

\subsubsection{Dataset transformations}
\label{sec:dataset-transform}

\label{h.dkl661gkybmp}
When tracking data flow, we are particularly looking for flows that are prone to leak information across multiple rows in a dataset or create new rows from old rows. 
The key insight is that transformations that remove rows or process data one row at a time are not problematic from a leakage perspective,
but transformations that process multiple rows together (as simple as counting or as complex as normalizing data by the largest value in a column) may cause leakage as rows may no longer be independent. 
In terms of typical concepts in functional programming, \textit{map}-like and \textit{filter}-like\footnote{
Technically, filter functions may change the length of a dataset and could leak information in the sense of non-interference. Pragmatically, we consider filter as unproblematic for data leakage, as filter is commonly used by data scientists for removing data points and therefore reporting this would not be very useful.}
transformations are okay, but \textit{reduce}-like transformations are suspicious.

To this end, our analysis labels 
(a) data-flow edges between datasets that perform computations independently on each row (map-like transformations) as \emph{map} edges and
(b) edges originating from datasets that perform computations across rows (reduce-like transformations) as \emph{reduce} edges. 
We manually label APIs in popular libraries (numpy, pandas, and sklearn) to identify those that perform computations per rows, 
such as \emph{fillna} (replaces missing values with the provided value) and those across multiple rows, 
such as \emph{numpy.mean} (computes a value based on multiple rows) and \emph{SelectPercentile.fit} (learns features from analyzing all rows). 
Notice that a \emph{reduce} edge may produce a single scalar value (such as calculating the mean of a column) which may be used subsequently in a map-like transformation (e.g., replacing missing values with that mean). 
Our analysis will later use this information to detect leaks when data flows through a \emph{reduce} edge.

In a similar fashion, we also identify transformations that duplicate rows (e.g., \emph{SMOTE.fit\_resample} from Fig.~\ref{code:overlap_example}) as \emph{duplicate} edges, 
which are potentially problematic because they can create dependencies between rows within a dataset, making random splits no longer produce independent datasets. 

In our running example (Fig.~4a), 
we see a preprocessing operation that may introduce leaks: \emph{SelectPercentile.fit}, 
as it uses the distribution information of the input data. It corresponds to a \emph{reduce} edge from \emph{X\_0} to \emph{select}.
Operation \emph{SelectPercentile.transform}, on the other hand, corresponds to a \emph{map} edge from \emph{X\_0} to \emph{X}.

\subsubsection{Related-data relations}
\label{sec:related-data-rel}

\begin{figure}[t]
\begin{subfigure}[b]{0.3\columnwidth}
  \centering
  \includegraphics[width=\columnwidth]{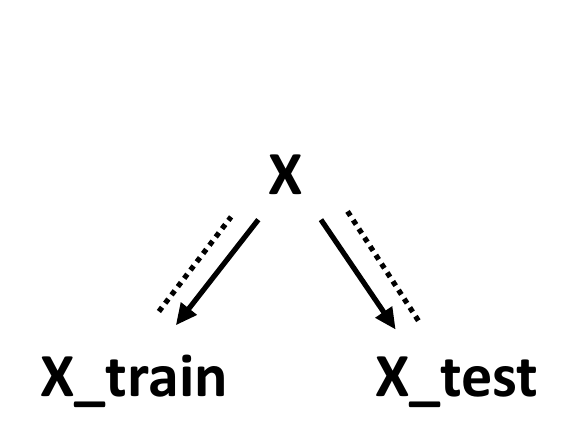}
  \caption{}
  \label{fig:rel-data:a}
\end{subfigure}
\begin{subfigure}[b]{0.3\columnwidth}
  \centering
  \includegraphics[width=\columnwidth]{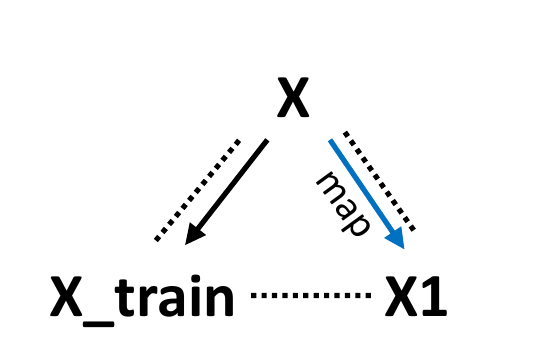}
  \caption{}
  \label{fig:rel-data:b}
\end{subfigure}
\begin{subfigure}[b]{0.3\columnwidth}
  \centering
  \includegraphics[width=\columnwidth]{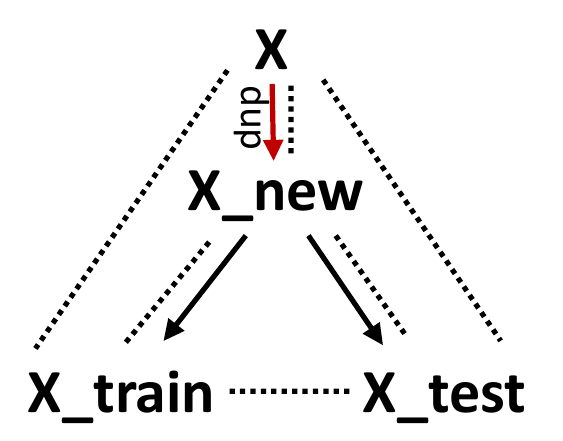}
  \caption{}
  \label{fig:rel-data:c}
\end{subfigure}
\vspace{-1em}
\caption{Related-data relations are usually not transitive (a), except for duplicate edges (b) or map edges (c). Solid arrow edges are normal data-flow edges; dotted edges are related-data relations.
}
  \label{fig:rel-data}
  \vspace{-1em}
\end{figure}

\looseness=-1
In many steps for detecting leakage, we need to identify whether two datasets are related or independent. 
For example, to reason about repeated evaluations with test data (multi-test leakage), 
we need to understand whether the data used in two test locations
are independent or somehow related. 
Technically, we establish a \emph{related-data relationship} to track whether two variables relate to each other (\textit{RelData} in Fig.~\ref{fig:rules}).
We consider two datasets as related (a) if they contain the same data, (b) if one is derived from the other, or (c) if they share rows from the same origin.

First, two datasets with a direct dataset flow (i.e., data-flow that only considers datasets) relation are considered as related (rule \textit{reldata/flow} in Fig.~\ref{fig:rules}). 
This is commonly the case when one dataset is a transformed version of another (e.g., Line 9 in our running example transforms \textit{X\_0} into \textit{X}) or simply assigned from one variable to another. 
Also, splitting a dataset creates multiple datasets that are each considered related to the original dataset (e.g., in Line 11 in our running example, \textit{X\_train} and \textit{X\_test} are both related to \textit{X}).
Our related-data relationship is reflexive and symmetric (rules \textit{reldata/ref} and \textit{reldata/sym} in Fig.~\ref{fig:rules}).

When two datasets are derived from the same original dataset, reasoning about their relationship is more complicated. 
In the common case that a dataset is split correctly into independent training and test sets, these two derived datasets should not be considered as related. 
However, when two datasets are derived from the same rows of an original source (e.g., two different ways of normalizing the entire dataset) they should be considered as related. 
Ideally, we would reason about transformations at the row level and consider datasets as related if their rows' origins in another dataset overlap, 
but corresponding dynamic analyses \cite{titian, newt, ramp} create runtime overhead and are difficult to adapt to our analysis. 
Instead, we approximate the most common patterns statically.

By default, we assume that two datasets derived from the same origin dataset are derived with a correct split and are hence considered as not related (Fig~\ref{fig:rel-data:a}). 
However, if we know that one of the datasets is derived with a map-like transformation (\textit{map edge} in Sec~\ref{sec:dataset-transform}) that dataset shares all rows with the original dataset and is not involved in a split. 
Hence, we consider that dataset as related to all other datasets derived from the same origin (Fig.~\ref{fig:rel-data:b}; rule \textit{reldata/map} in Fig.~\ref{fig:rules}). 
Furthermore, if the origin dataset contains dependent rows (e.g., from a data augmentation step), we assume that derived datasets are also related, as splits are unlikely to maintain independence. 
Hence, we consider all datasets derived from an origin where data was produced (directly or indirectly) through a duplicate edge (see Sec~\ref{sec:dataset-transform}) as related (Fig~\ref{fig:rel-data:c}; rule \textit{reldata/dup} in Fig.~\ref{fig:rules}). 
\looseness=-1

Note that our three heuristics approximate a more accurate dynamic analysis, but cannot cover all cases. 
For example, our heuristic would identify a split with overlapping rows (\textit{a, b = o[:100], o[50:]}) as independent, but those mistakes are very rare. 
We consider that the cost of increased false positives or more expensive dynamic analysis outweigh the benefits of detecting such very rare cases. 
As we will show in our evaluation in Section~\ref{sec:empirical-study}, our heuristics capture the relationships found in common notebooks to achieve relatively accurate leakage detection.

\subsection{Data-Model Mappings}
\label{sec:model-data}

While data-flow analysis tracks how data is transformed and moves through the notebook, 
we can only identify whether data is training data, validation data, or test data by determining how it is eventually used.

As a first step, we identify program locations where (usually preprocessed) data is used for training, validation, and testing. 
We identify those locations simply by finding API calls in the notebook that are typically used for training and evaluation purposes, such as the \textit{fit} and \textit{predict} functions in \textit{sklearn}'s APIs.
We later trace back the source of the data used in these APIs to processing steps and original datasets using standard data-flow analysis.

\looseness=-1
While training data can be identified with distinct function calls, distinguishing between validation and test data is conceptually challenging because both are used with the same APIs (e.g., predict). 
We cannot reliably infer whether a data scientist intends to use data for validation or testing purely based on notebook code---this is a challenge even for human experts who may need to rely on context clues or documentation. 
We therefore rely on a simple heuristic that considers data that is used repeatedly in evaluation as validation data and all data that is used only once in evaluation as test data.
We consider data to be evaluated repeatedly if its location is within a loop or if two locations are connected to evaluate the same or related data as per our data-flow analysis (i.e., variables in both locations are connected through the related-data relation from Sec.~\ref{sec:related-data-rel}).

Finally, we group each training dataset with the corresponding validation and test datasets that relate to the same model. 
We use standard data-flow analysis to identify which call locations share the same target object. To account for possible repeated training of the same model object, 
we always group training data with all subsequent validation and test data, until the next training data is identified. 
In the end, we derive a series of model-data tuples (\textit{ModelData} in Fig.~\ref{fig:rules}),
where the same training data might correspond to zero or multiple validation/test datasets. 
In our example (Fig.~4a), 
we show two model-data tuples based on this heuristic: (\textit{lr}, \textit{X\_train}, \{\textit{X\_test}\}, $\varnothing$) and (\textit{ridge}, \textit{X}, \{\textit{X\_test}\}, $\varnothing$).

\subsection{Leakage Detection}
After collecting the above information, identifying leakage is performed through pattern matching:

\begin{itemize}\item
\looseness=-1
To detect multi-test leakage, we check for a given model whether there exists at least one piece of test data. 
Note that `test data' evaluated multiple times will already be identified as validation data in our analysis (see Sec.~\ref{sec:model-data}). 
If for all models, there is no test data but only validation data detected, we report multi-test leakage (rule \textit{leak/multi} in Fig.~\ref{fig:rules}).

\item
To detect overlap leakage, we check for a given model \linebreak whether training data and test/validation data\footnote{
Note that for overlap/preprocessing leakage, we do not distinguish between test and validation data. Leakage between training data and validation data is still problematic, as it defeats the purpose of validation data (i.e., providing independent validation during model development).}
are related. 
Note that there might be multiple test/validation data in a single model-data tuple (see Sec.~\ref{sec:model-data} for how we derive the tuple). 
Therefore, for a given trained model, we only report overlap leakage when all of its test/validation data overlaps with the training data  (rule \textit{leak/overlap} in Fig.~\ref{fig:rules}).

\item
To detect preprocessing leakage, we check whether training data contains information from test/validation data through preprocessing (\emph{reduce} edges). 
If training data uses reduced information from test/validation data (or datasets that are related to test/validation data), we will report a case of preprocessing leakage  (rule \textit{leak/pre} in Fig.~\ref{fig:rules}).

\end{itemize}

In our example, we could find a path from \textit{X} to \textit{X\_train} and a path from \textit{X\_0} to \textit{X\_test} (see Fig.~4c). 
As \textit{X} contains reduced information from \textit{X\_0}, which is related to \textit{X\_test}, 
we establish that test data information is leaked into training data, and there is a preprocessing leakage.
Next, because \textit{X} is related to \textit{X\_test} (as \textit{X\_test} is transformed from \textit{X}), 
there is an overlap leakage when we evaluate the second model \textit{ridge}.
Finally, the two trained models share the same test data (\textit{X\_test}), which we will identify as validation data. 
Because there is no independent test data used in the final evaluation, we conclude that there is also a multi-test leakage.

In the fixed version of our example (see Fig.~4b), 
we see that the two model-data tuples are changed to (\textit{lr}, \textit{X\_train}, \{\textit{X\_test}\}, \{\textit{X\_test\_new}\}) and $(\textit{ridge}, \textit{X}, \{\textit{X\_test}\}, \{\textit{X\_test\_new}\})$ and hence: (1) models no longer contain information from \textit{X\_test}, as the reduced information only comes from \textit{X\_train} (see Fig.~4d)), 
 eliminating the preprocessing leakage, (2) in all tuples, training data and test/validation data are no longer related, eliminating the overlap leakage, and (3) there is independent test data \textit{X\_test\_new} that is evaluated only once, eliminating the multi-test leakage.

\subsection{Implementation}
To make our analysis easy to extend and modify, our implementation uses \textit{datalog}, a language commonly used in declarative program analysis~\cite{smaragdakis2015pointer, bravenboer2009strictly}. 
Our two-phase implementation first transforms Python code into datalog facts as an intermediate representation and then analyzes these facts to generate leakage detection results. Our analysis design is similar to \emph{doop}~\cite{bravenboer2009strictly}, a popular Java program analysis framework.

In the front end, we generate datalog facts that can be easily analyzed subsequently. Specifically, we
translate complex language structures into simpler ones,
translate assignments (both variables and fields) to static single assignment form, which ensures that subsequent analyses are flow-sensitive,
and match method invocations with signatures.
To identify datasets and targets of method invocations, we perform type inference with the off-the-shelf type inference engine \textit{pyright}~\cite{pyright}.

Based on initial datalog facts, we compute additional facts for the relations (e.g., \emph{RelData}, \emph{ModelData}) described above and subsequently detect data leakage using datalog queries.
We implement a standard Anderson-style 2-call-site-sensitive pointer analysis similar to \emph{doop}, with special treatment of common language features (e.g., lists and global variables). 
The data-flow analysis is built on pointer analysis and also follows standard implementations.

Our analysis requires specifications of data science APIs. Specifications are mainly used to provide domain knowledge (e.g., which APIs are used for training/testing, which APIs behave \textit{reduce}-like).
Our current implementation supports three machine learning libraries -- \textit{sklearn}, \textit{keras}, and \textit{pytorch} -- and two libraries commonly used for data transformations -- \textit{pandas} and \textit{numpy}. 
We went through the official documentation of these libraries to find APIs that perform data transformations and
APIs that perform supervised learning. 
Our analysis can be easily extended to other libraries by providing their specifications.

\section{Evaluation}
We first evaluate the accuracy of our analysis, that is, its ability to find actual leakage and to avoid false alarms:
\begin{itemize}
    \item \textbf{RQ1}: \textit{How accurate are the results of our analysis?} 
\end{itemize}

To ensure that our analysis can be used in an interactive or continuous integration setting during  model development, we also evaluate  efficiency in terms of running time:
\begin{itemize}
    \item \textbf{RQ2}: \textit{How efficient is our analysis?} 
\end{itemize}

Finally, after establishing accuracy and efficiency, we use our analysis to study test-train leakage in a large corpus of notebooks, exploring common forms and sources of leakage. We will show that leakage is common across different types of notebooks. In addition, leakage often manifests itself in nontrivial data flows in notebooks, in forms that can be tedious or even difficult to detect manually, providing strong, albeit indirect evidence for the usefulness of our automated detection:
\begin{itemize}
    \item \textbf{RQ3}: \textit{How prevalent is data leakage in public notebooks?}
    \item \textbf{RQ4}: \textit{What do typical leakage issues look like?}
\end{itemize}

\subsection{Research Design}
We evaluate all four research questions with a corpus of public data science code in Jupyter notebooks. For different research questions, we use different subsets of this corpus.

\subsubsection*{Corpus of notebooks with data science code.}
To answer our research questions, we curate a large corpus of public notebooks with data science code. 
Specifically, we collect Jupyter notebooks from GitHub and Kaggle.
GitHub is a common platform for storing data science code for a range of purposes, from hobby and educational projects, to research projects and tutorials, to production systems. Kaggle is a common platform for data science competitions where users can submit notebooks as solutions to competition problems.
We purposely selected code in notebooks, rather than arbitrary Python files, 
because notebooks are the primary environment for developing data science code~\cite{perkel2018jupyter}.

For GitHub, we collected \emph{all} notebooks from GitHub repositories created in September 2021 (strictly independent from all notebooks from earlier periods used during development of our analysis). Specifically, we used the GitHub search API to identify repositories with notebook code and partitioned the search space to collect \emph{all} \numprint{81026}~repositories. By selecting all notebooks from a recent time period, we get a full and representative sample of the different kinds of notebooks published on GitHub. We collected a total of \numprint{280994}~notebooks this way.

For Kaggle, we selected a smaller and more targeted notebook population, 
collecting notebooks from two popular competitions, \textit{titanic} and \textit{housing} \cite{titanic, housing}. 
For each competition, we collect the 200 notebooks with the most votes and the 200 most recent notebooks, as of April 12, 2022.
We selected these competitions, because they use tabular data and require significant preprocessing effort. 
We use this corpus to understand leakage issues among typical competition solutions that are prone to data leakage. 
They are not necessarily representative of all competition solutions on Kaggle.

We further filter these notebooks to include only those that use the machine learning libraries supported by our current implementation (\emph{sklearn}, \emph{keras}, and \emph{pytorch}). 
The discarded notebooks either only use not-yet-supported libraries such as \emph{tensorflow} or do not train any models. This leaves us with \numprint{107603} GitHub notebooks and \numprint{108273} in our corpus overall. In Table~\ref{tab:leakage}, we show descriptive statistics of our final corpus.

\begin{table*}[t]
\small
\begin{tabular}{lllllllll}
\toprule
Dataset & \#notebooks & LoC & \#stars & Preprocessing Leakage & Overlap Leakage & Multi-test Leakage & Any Leakage  \\
\midrule
GitHub (all)  & \numprint{107603} & 410  & 1.1 & 12.3 & 6.5 & 18.5 & 29.6 \\
GitHub (popular) & \numprint{920} & 378  & 95.2 & 4.9 & 5.2 & 15.9 & 20.9 \\
GitHub (tutorials) & \numprint{1157} & 584  & 4.2 & 3.9 & 2.9 & 11.3 & 16.2 \\
GitHub (assignments)& \numprint{7576} & 559  & 0.6 & 13.9 & 7.4 & 22.0 & 33.0 \\
Kaggle (top)  & 312 & 851 & - & 56.1 & N/A & N/A & N/A \\
Kaggle (recent)& 358 & 504 & - & 55.8  & N/A & N/A & N/A \\
\bottomrule
\end{tabular}
\caption{\small Data Leakage Distribution. LoC is the average number of lines of code across all notebooks in the group.
Number of stars is based on the repository that these notebooks reside in and is also averaged. 
We show the percentage of notebooks for which we report each leakage type. For Kaggle notebooks, we only track preprocessing leakage because the other two are infeasible in this setting (marked as N/A in the table).}
\label{tab:leakage}
\vspace{-1.5em}
\end{table*}

\subsubsection*{Analyzing accuracy (RQ1).}
Establishing ground truth for data leakage is challenging and we are not aware of existing datasets.
To evaluate the accuracy of our analysis, we measure both false positives and false negatives on a sample of notebooks for which we manually establish ground truth. 
Due to the substantial manual effort involved, we perform the analysis on 100 randomly sampled notebooks from our GitHub corpus (which yields an 8\,\% error margin with a confidence level of 95\,\%)~\cite{samplesize}.

One author manually labeled these 100 notebooks looking for leakage issues and then compared manual labels with the analysis results.
For all notebooks where the manual labels and analysis results disagreed,
the author sought the expert opinion of a second author (a trained data scientist). Together they discussed the issue to determine whether the notebook contained leakage, correcting the ground truth label if needed. 
This correction was needed in 7 notebooks, which were incorrectly labeled initially, arguably indicating that manual checking of data leakage is
non-trivial and error-prone, even for experts. 
Overall our process balances labeling effort and confidence in the ground truth.

\looseness=-1
As per our manually established ground truth, 40 of the 100 notebooks contained at least one form of leakage -- 20 with preprocessing leakage, 8 with overlap leakage, and 32 with multi-test leakage.

\subsubsection*{Analyzing efficiency (RQ2).}
For efficiency, we recorded the execution time of our analysis for \emph{all} notebooks in our GitHub corpus. We record timing separately for the analysis front end (collecting facts), the type inference, and the actual analysis (datalog engine). We set a timeout of 5 minutes per notebook.
The experiment is conducted on a Precision 3650 workstation, with Intel(R) Xeon(R) W-1350 CPU and 32GB memory. The time is measured using Python's time module (wallclock time).

\subsubsection*{Analyzing leakage frequency (RQ3).}
We analyze leakage for the entire corpus and report the frequency with which we raise warnings for each kind of leakage. Note that we consider at most one warning per leakage kind per notebook to avoid biasing results with some notebooks that raise lots of warnings.

To further understand whether leakage associates with certain types of notebooks, we separately report leakage for different subpopulations. Specifically, we break down results for the following subpopulations:
\begin{itemize}
\item \emph{Popular notebooks} are viewed (and possibly reused) by more people, thus having more potential to spread problematic practices. We conjecture that popular notebooks come from more experienced data scientists and are better crafted. We identify 920 such popular notebooks as those in GitHub repositories in our corpus with 10 or more stars. 
\looseness=-1
\item \emph{Tutorial notebooks} similarly are explicitly designed for teaching others and could spread problematic practices. We identify \numprint{1157} tutorial notebooks by searching for the phrase \textit{``this tutorial''}. 
\looseness=-1
\item \emph{Assignment notebooks} contain solutions to course projects and assignments. We conjecture that these notebooks better represent practices of beginners than average notebooks in our corpus.
We identify \numprint{7576} assignment notebooks on GitHub by  searching for the keywords \textit{`homework'} and \textit{`assignment.'}
\item \emph{Competition notebooks (popular and recent)} are written by a mix of experienced and learning data scientists, possibly with an increased incentive to maximize model accuracy. Here, we report results from the Kaggle notebooks from our corpus.
\end{itemize}

\subsubsection*{Analyzing leakage characteristics (RQ4).}
We measure distance between different program constructs by measuring lines of code between them (based on the Python files converted from notebooks using \textit{nbconvert} in the default setting) and compare them to the length of the entire notebook.
We expect that issues that span longer distances are harder to analyze manually.
Specifically, we calculate distance between leakage sources (e.g., reduced data) and training locations for preprocessing leakage, 
and distance between different evaluation locations for multi-test leakage.
We report the results for the entire GitHub corpus. 

For the whole dataset and each sub-population, we explore the distributions of different leakage issues, complexities of these issues, 
and also how they are distributed across different sub-populations.

\subsubsection*{Threats to validity.}
Establishing accurate ground truth for leakage is challenging. Our experience shows that even data science experts looking for leakage may miss it in complex data flows. We adopt a best effort approach with human labeling and comparisons with automated results that balance effort with confidence. We share our data for independent validation.

For RQ3 and RQ4, we report leakage warnings but validating all warnings is simply infeasible at this scale. Our results should therefore be interpreted with the error margins established in RQ1.
In addition, warnings about multi-test leakage may be rooted in settings where independent test data may exist outside of the notebook; our evaluation would also not detect multi-test leakage if the same test data was used repeatedly in past versions of a notebook or is used repeatedly in multiple notebooks. Generally, our analysis provides only a piece of a larger picture that needs to involve process design and other assurances, as we will discuss in Section~\ref{sec:discussion}.

The notebook population in our corpus is representative of public notebooks on GitHub (and some Kaggle competitions), but may not generalize to data science code outside of notebooks, across multiple notebooks, or to proprietary data science pipelines. Readers should hence be careful when generalizing our results.

Finally, our analysis focuses on the presence of leakage, not whether data scientists find leakage reports actionable or what effect leakage has in overestimating the reported accuracy results. 
We leave such evaluations to future work but point again to the fact that leakage is firmly established as problematic in educational material (cf. Sections~\ref{sec:intro}--\ref{sec:overfitting}).

\subsection{Results}

\subsubsection{Analysis accuracy (RQ1)}
Our analysis prototype successfully executed for 94 of the 100 notebooks in our labeled RQ1 sample; the remaining 6 notebooks failed due to syntax errors. For the 94 notebooks, our prototype found 15 with preprocessing leakage, 7 with overlap leakage, and 19 with multi-test leakage.
All the found leakage issues were true positives except for one case due to over-approximation in related-data analysis, yielding a precision of 97.6\% (40 out of 41 detected issues).

\looseness=-1
On the other hand, our prototype missed 19 issues due to unsupported libraries~(6),
mistaking single test cases as test data (3), storing/loading model in external storage~(3), undetected test data evaluation (3), 
inaccurate type inference (2), and under/over-approximation in related-data analysis (2). This yields a recall of 67.8\% (40 out of 59).

\emph{Overall, our prototype achieves an accuracy of 92.9\% (262 out of 282 potential leakages).}
Analysis for preprocessing/overlap leakage is more accurate than multi-test leakage in this sample. 
Based on these results, we conclude that our analysis is generally accurate. 


\subsubsection{Analysis overhead (RQ2)}



Most (92.78\,\%) of the 107,603 GitHub notebooks in our corpus could be analyzed successfully within the 5 minute time limit.
On average, our analysis completes within 3.23 seconds, with most of the time (2.20 seconds on average) spent on type inference.
A small percentage of notebooks could not be analyzed due to
syntax errors (7.08\,\%), timeout (0.09\,\%; usually due to explosion from context-sensitivity in our data-flow analysis), and language features not supported by the front-end parser (0.05\,\%, e.g., named expression introduced in PEP 572).
\emph{We conclude that our analysis is efficient enough for interactive use and in continuous integration settings.}


\subsubsection{Leakage in public notebooks (RQ3)}
\label{sec:empirical-study}

Our analysis reports at least one form of leakage for almost a third of all public notebooks in our GitHub corpus (29.6\%, see Table~\ref{tab:leakage}). \emph{We found frequent evidence of all three forms of leakage.}

\emph{Preprocessing leakage is prevalent in notebooks.}
Overall, our analysis reported preprocessing leakage for 12.3\% of notebooks in our corpus.
The most common sources of leakage during preprocessing are scaling, computing mean and standard deviation, and using results of a principal component analysis (PCA) in downstream data transformations.
For text data, the most common source of preprocessing leakage is vectorizing through counting or \emph{tf-idf} over the whole dataset. 
Zooming in, we reported preprocessing leakage in 32.9\% of those notebooks that scale their data, 8.4\% of those that compute mean or standard deviation, and 13.4\% of those that perform PCA.

\emph{Many notebooks lack independent test data.}
We report multi-test leakage in 18.5\% of all notebooks in our corpus. 
We also detected that among all notebooks that train a model, 53.9\% contain validation data
(i.e., data that is used repeatedly for evaluation), but 35.0\% of trained models are not evaluated with \textit{independent} test data. 
Models evaluated with validation but without test data represent 28.3\% of all trained models in our GitHub corpus.

\emph{Training data often overlaps with validation and test data.}
We report overlap leakage in 6.5\% of all notebooks in our corpus. 
A closer look reveals that 8.0\% of all trained models are only evaluated on data that overlaps with its training data.


\emph{Leakage is common in both beginner and expert code.}
When analyzing the subsets of our corpus, we report leakage in all subsets,
but to different degrees (see Table~\ref{tab:leakage}).
Assignment notebooks more representative of beginners are more likely to receive reports for all leakage types. In contrast, popular notebooks and tutorials
most likely associated with more advanced data scientists are slightly 
less likely to receive leakage reports, but leakage is still reported fairly commonly (20.9\,\% of popular and 16.2\,\% of tutorial notebooks). The fact that leakage seems common even in tutorial notebooks designed for educational purposes seems concerning.

Finally, the rate of reports of preprocessing leakage is very high in competition notebooks (>55\,\%). 
Indeed, it is not uncommon that competitors concatenate separately provided train and test data before preprocessing the combined dataset.
The way the competitions are designed (providing values but not labels of test data) may encourage exploiting leakage to maximize accuracy results, even, or especially by, experts. 
At the same time, because test labels are not provided, we do not report overlap and multi-test leakage. We will discuss competitions separately in Section~\ref{sec:discussion-kaggle}.



\subsubsection{Leakage characteristics (RQ4)}
We observe that leakage often occurs in patterns that make it challenging to detect manually.

\begin{figure}[t]
\begin{subfigure}[t]{0.45\columnwidth}
    \centering
  \includegraphics[width=\columnwidth]{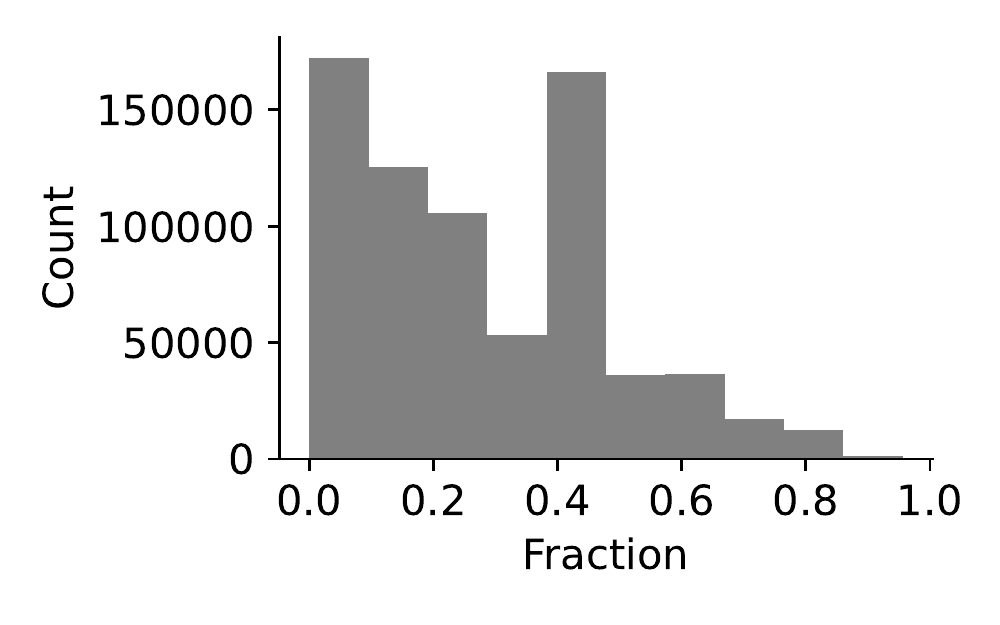}
  \caption{\small Distribution of fraction of distance between  preprocessing leak sources and training data.}
  \label{fig:frac1}
\end{subfigure}
\quad
\begin{subfigure}[t]{0.45\columnwidth}
  \centering
  \includegraphics[width=\columnwidth]{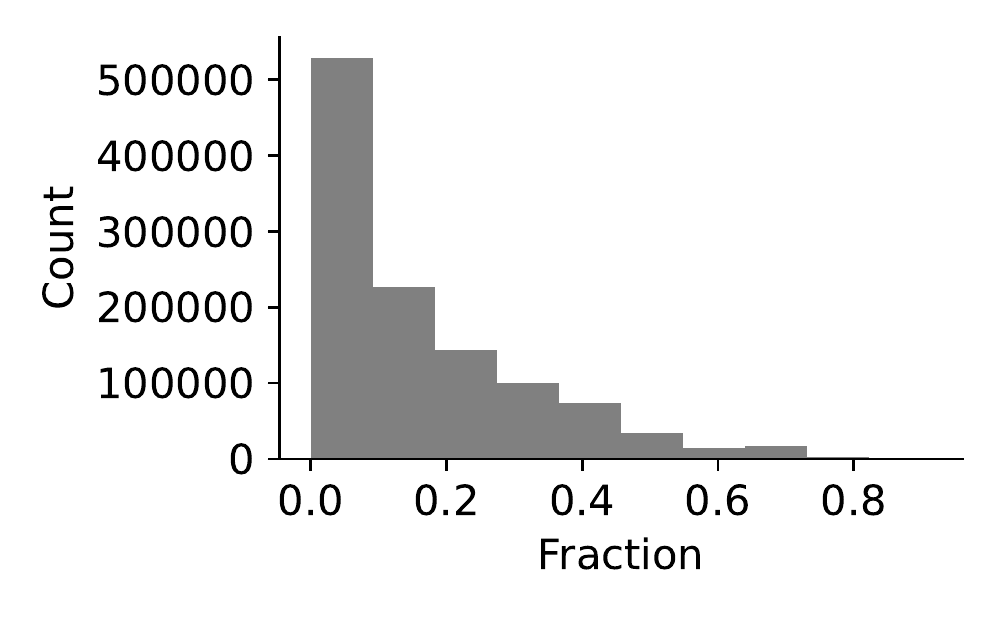}
  \caption{\small Distribution of fraction of distance between test locations that evaluate the same data.}
  \label{fig:frac2} 
\end{subfigure}
\vspace{-1em}
\caption{Leakage issues exhibit non-local pattern.}
\label{fig:frac}
\vspace{-2em}
\end{figure}

\emph{Leakage issues exhibit non-local patterns.}
For preprocessing leakage, the average distance between the leakage source (the \textit{reduce} edge) and the location where the training data is used is 293~lines of code.
In more than half of the cases, the distance between leakage source and training location is more than 20\% of the length of the notebook (see Fig.~\ref{fig:frac1}).
This distance illustrates the often long processing sequences and non-local data flows that are difficult to analyze manually.


\looseness=-1
For multi-test leakage, the average distance between two locations evaluating models with the same (or related) data is 255~lines of code.
In more than 30\% of all cases, test location distance is more than 20\% of the whole notebook (see Fig.~\ref{fig:frac2}).
On average, there are 4.4 model-evaluation locations that use the same (or related) test data in notebooks with a multi-test leakage warning. This similarly illustrates the non-local reasoning required to notice this form of leakage.



\emph{A single notebook often trains multiple models.}
Among all notebooks in our GitHub corpus, 65.3\% train at least one model (in sklearn, pytorch, or keras) and 66.0\% evaluate at least one model (5.8\% of notebooks do not train but evaluate a model, typically when loading a pre-trained model).
Among the notebooks that do train at least one model, we found that 54.3\% train multiple models.
Having to commonly track multiple models and how data flows into their training and evaluation can be another challenge when manually reasoning about leakage.










\section{Discussion}
\label{sec:discussion}

Our results indicate that static detection of several forms of data leakage is feasible and that this kind of leakage is pervasive in practice. 
At the same time, it is not a comprehensive solution to avoid leakage or other forms of overfitting.




\subsection{Practical Impact of Data Leakage}
Not all leakage issues are equally problematic and some data scientists developing models may consider that some forms of leakage (e.g., the median of a column) are entirely negligible and that therefore warnings about leakage are not actionable and hence \emph{effective false positives}~\cite{tricorder}.
We even found some tutorials where the description explicitly indicates that developers are aware of leakage problems but ignore them anyway, such as 
testing with part of the training data ``for the sake of simplicity''.


We have seen and heard of a large range of different impacts. On the one hand, we have heard (personal communication) of multiple cases where models were accidentally evaluated on training data producing entirely misleading results in research teams at BigTech companies, and that it is easy to create artificial examples where preprocessing leakage creates substantially inflated accuracy results (e.g., Fig.~\ref{fig:leak_acc}). On the other hand, experiments with notebooks in our corpus often just yielded marginal if any differences in accuracy.\footnote{Due to the known problems of reproducing public notebooks~\cite{pimentel2019large, wang2020assessing}, we were not able to perform systematic experiments on a larger sample.}

We expect that more substantial leakage, such as evaluating on the training data mostly stems from simple mistakes (such as using the wrong variable name) -- which \emph{practitioners} can easily detect with our analysis. The more subtle leakage through preprocessing may often have little effect on reported accuracy in most cases, but we still argue that it represents a bad coding pattern.
In our evaluation, we found that even many tutorials and top competition solutions leak their test data, let alone homework assignments.
We think that in particular \emph{educators} should insist on avoiding leakage in their course materials and homework, and our analysis provides an easy way to create awareness of the most common patterns leading to leakage.\footnote{We see some practitioners and educators agree with this stance. For example, in this GitHub issue (\url{https://github.com/keras-team/keras/issues/1753}), participants actively discuss the potential problems of
tutorials setting a bad example for learners, even when there technically is no actual leakage problem in the specific example. 
}



\subsection{Process Design for Preventing Data Leakage}
While we intend our analyzer to be used primarily as an educational tool and a tool to surface common mistakes in practice (either directly in notebook environments or integrated into code reviews~\cite{tricorder}), a more robust solution can be achieved by designing the process of how responsibilities are assigned. This is particularly relevant in formal settings where model development is outsourced~\cite{nadiaICSE22} and in data science competitions.

\subsubsection{Contract settings.}
In settings where a team is given data to develop a model as part of a contract (e.g., outsourcing), data leakage can systematically be avoided if test data is not provided to the data scientists who develop the model in the first place, but instead reserved for external evaluation upon delivery. 
When data scientists do not have access to test data, they cannot derive any insights from it, cannot use it repeatedly in evaluations, and cannot even manually look at data distributions to inform modeling decisions---even in settings where a data science team would have an incentive to cheat to present inflated accuracy numbers. 
The drawback of such a process design is that an additional external evaluator is needed who must have enough understanding of the data to be able to split it appropriately into training and test data (which can be nontrivial when dependencies exist~\cite{gorman-bedrick-2019-need, sogaard-etal-2021-need, splits-aiops,Saeb2017TheNT}). 
The evaluator also needs to have access to the model to run it locally, to not risk leaking test data during model inference. 
Also, the evaluator cannot repeatedly report results from the evaluation back to the developers of the model. Notice that test data does not become immediately useless, but gradually loses confidence which can be accounted for in scores, as explored in detail elsewhere \cite{renggli2019mlsys,thresholdout}.

\looseness=-1
In many practical settings though, the team developing the model is involved in acquiring or collecting the data in the first place~\cite{nadiaICSE22}. 
In such settings, it would be paramount for an external evaluator to independently collect data, which is often infeasible or prohibitively costly, as it might require replicating the expertise gained by the model development team during the development process. If there is some trust relationship between model developers and model users, approaches that foster best practices and avoid common mistakes, such as our static analysis tool, might be a better pragmatic alternative.



\subsubsection{Data science competitions.}
\label{sec:discussion-kaggle}
Competitions often pursue a similar form of external evaluation, but often make compromises to better automate the competition, reduce cost and complexity, or increase engagement with more granular feedback:

{\emph{Withholding test labels.}} A common design for Kaggle competitions is to provide test data without labels. The competitors submit the predicted labels to receive a test score. The operational advantage of this approach is that the competition organizer does not need to execute the submitted models---it avoids (1) executing submitted (untrusted) code, (2) having to support a range of different models, and (3) bearing the cost of model inference during evaluation. While this design prevents overlap leakage and multi-test leakage, preprocessing leakage is still possible. In fact, our evaluation of the two Kaggle competitions, which both use this design, shows that competitors often exploit preprocessing leakage.

{\emph{Limiting repeated submissions.}} Most competitions allow participants to submit multiple revisions of a solution, receiving scores for each of them, risking multi-test leakage. To minimize this risk, some competitions limit the number of submissions per team. However, if participants can see other solutions in a leaderboard or create multiple accounts, leakage is still a concern even if the actual test data is withheld from competitors.

\looseness=-1
{\emph{Partial test scores.}} Some competitions split their hidden test data and provide feedback on incremental submissions only with part of the test data, until finally scoring all submissions exactly once at the end of the competition with the remainder of yet unused test data. This effectively separates the hidden evaluation data into validation and test data. 
It still allows to provide meaningful preliminary feedback on a leaderboard during the competition, without allowing overfitting in the final results. As a downside, competitors that are better at extracting insights from partial test scores may have an advantage in the competition over those that just work with the training data, hence encouraging explicit attempts to leak information.

While a once-only external evaluation with hidden test data at the end of the competition may be the cleanest solution, in practice competition designers will often want to make compromises with one of the designs above, which each reduce leakage to some extent but cannot avoid it entirely. Adding our static analyzer to the competition infrastructure could call out bad practices or could remind observers to not use the same practices outside competitions.



\subsection{API Design for Preventing Data Leakage}
It is also possible to avoid certain leakage issues through API design. For example, many ML libraries advocate the use of pipelines for preprocessing~\cite{scikit}. If correctly used, they could help avoid preprocessing leakage, because the library enforces that only training data is used to extract parameters for preprocessing.

From our GitHub corpus, we observe that only 5.5\% notebooks use pipelines. 
Surprisingly, 18.1\% of these notebooks still contain pre-processing leakage. 
When we inspected samples from these notebooks, we found that they often do not apply pipelines to the whole preprocessing stage, and some of them even use pipelines in the wrong way.
This informs us that better education on how to use these APIs is as important as designing these leakage-proof APIs themselves.

\subsection{Limitations and Alternatives}
Our definitions of leakage and corresponding analyses are limited to the scope of what is observable statically in data science code (typically in a notebook). In addition, our static approximation relies on models of library functions and several heuristics.

\looseness=-1
Importantly, our approach cannot detect repeated evaluations that are not present in the notebook (e.g., cells modified and evaluated repeatedly) and cannot detect test data outside the notebook. It may therefore issue spurious  multi-test leakage warnings or miss some. 
Our evaluation also shows that our analysis misses some leakage but produces few false positives (recall: 67.8\%, precision 97.6\%), therefore the reported leakage may be underestimated.
Similar to many static analysis warnings, we envision warnings as a starting point for reflection and discussion (e.g., during code review~\cite{tricorder}), and not necessarily as a blocking issue that always needs to be addressed with code changes.
For example, developers might compare several techniques on the same dataset, while the final test score is computed outside the notebook. Our approach would report such patterns, but these false positives should not affect the developer's workflow.

In an adversarial setting, it is easy to trick our analysis by for example using meta-programming or coding patterns that exceed the capabilities of our analysis (e.g., store and load data in a file). 
Additional rules and environment models can strengthen our analysis but will not overcome fundamental limitations. 
A sound analysis for an adversarial setting may be possible but would likely be so restrictive or create so many false positives to render the approach impractical.
Again, our analysis is not a safeguard against all kinds of cheating in data science competitions. It better serves as a lightweight checker that discourages bad practices of data leakage.

Dynamic analysis could improve accuracy in many cases, for example tracking the origin of individual rows in the data rather than our approximations in the related-data relationship. However, dynamic analysis would require the entire notebook to be executed for analysis (with the induced overhead), which can be costly in many machine learning tasks and may not be feasible when studying public notebooks that are often hard to reproduce~\cite{pimentel2019large}.



\section{Related Work}

\paragraph{Quality Assurance for Data Science Code}
Prior work has noted that data science code is often of low quality---%
relying heavily on copied code and code clones~\cite{koenzen2020code},
ignoring basic coding and style conventions~\cite{zeller-icse20-nier},
being poorly documented~\cite{painpointspaper}, 
containing frequent bugs in data transformations~\cite{chenyangASE21},
and being hard to reproduce~\cite{pimentel2019large,wang2020assessing}.
In a well-known article, \citet{Sculley2015HiddenTD} have argued that data science code is particularly prone to accumulating technical debt due to complexity and often poor engineering practices.
Our work on data leakage explores another common quality issue that may lead to unreliable accuracy evaluations.


\paragraph{Static Analysis for Python}
Despite its popularity, there are relatively few static analysis frameworks or tools written for Python,
partly due to the difficulties of handling Python's dynamic features.
\citet{head2019managing} implemented a static def-use analysis for Python to perform program slicing that helps data scientists clean, recover, and compare versions of code.
\textit{Scapel}~\cite{li2022scalpel} is a static analysis framework written in Python that integrates several common analyses (e.g., alias analysis).
\textit{NBLyzer}~\cite{DBLP:journals/corr/abs-2110-08339} is a framework specifically written for notebook code in Python, 
where they focus on supporting notebook actions (e.g., code changes, cell executions).
Pysa~\cite{pyre} is a taint analysis tool that aims to identify potential security issues in Python code.
~\citet{tensorflowshape20} proposed a static shape analysis for Tensorflow programs, which is integrated into the \textit{doop} framework.
We chose to develop our own analysis, because it gives us the flexibility to tailor it for the purpose of leakage detection,
where we need to track several custom relations.


\paragraph{Data Leakage Detection}
\looseness=-1
Data leakage detection is a largely unexplored problem.
\citet{kaufman2012leakage} discuss how to manually perform analyses to detect certain kinds of data leakage in raw data, in particular label leakage.
\textit{Deepchecks}~\cite{deepchecks} is a library that aims to validate machine-learned models and data, which
supports dynamic check for overlap leakage between train and validation/test set by dynamically inspecting datasets (e.g., whether test data contains rows that occur identically in training data). 
Closest to our work is a customized analysis to detect preprocessing leakage in the \textit{NBLyzer}~\cite{DBLP:journals/corr/abs-2110-08339} framework.
However, their data leakage analysis is just a demonstration for their framework and does not capture the full complexity of data leakage.
For example, they do not establish data-model mappings but assume that all training/test locations are relevant globally (i.e., connected by the same/related models).
They also do not actively distinguish validation data from test data, or track whether variables might be aliased.
This results in a leakage specification that only poorly approximates the ground truth.
Because their implementation is not publicly available, we did not compare our approach with theirs in our evaluation.
In contrast to prior work, we detect multiple forms of data leakage and also detect leakage accurately even when datasets are transformed.


\paragraph{Provenance Tracking for Data Science Code}
Provenance tracking in data science code has been studied extensively~\cite{Vamsa, pimentel2017noworkflow}.
Most relevant are approaches that track origins of data at the row level in data analytics code for various frameworks: Titian~\cite{titian} for Spark, RAMP~\cite{ramp} for Hadoop, and Newt~\cite{newt} for Hadoop and Hyracks.
As discussed, we only statically approximate what could be tracked more accurately with dynamic record-level provenance tracking approaches, but these approaches need to instrument the frameworks to track provenance at runtime.
Overall, this kind of tracking is only a building block in our leakage detection.



\section{Conclusion}


\looseness=-1
We provide a summary of common data leakage problems and propose a static analysis approach that could automatically detect them.
We find that leakage issues are common in public notebooks and provide recommendations on process designs to prevent data leakage.


\begin{acks}
Kästner and Yang’s work was supported in part by NSF awards 1813598 and 2131477. 
Lewis and Brower-Sinning's work was funded and supported by the Department of Defense under Contract No. FA8702-15-D-0002 with Carnegie Mellon University for the operation of the Software Engineering Institute, a federally funded research and development center (DM22-0666).
\end{acks}


\balance
\bibliographystyle{ACM-Reference-Format}




\appendix

\end{document}